\documentclass[10pt]{iopart}

\usepackage{amssymb}
\usepackage{amsfonts}
\usepackage{float}
\usepackage[caption = false]{subfig}
\usepackage{graphicx}    
\usepackage{iopams}  
\usepackage{cite}

\graphicspath{{./}{figures/}} 

\begin{document}

\title{Multi-Ciliated Microswimmers -- Metachronal Coordination and Helical Swimming}

\author{\noindent\large{Sebastian Rode, Jens Elgeti, and Gerhard Gompper} \\}

\address{Theoretical Physics of Living Matter, 
	Institute of Biological Information Processing and Institute for Advanced Simulation, 
	Forschungszentrum J\"ulich, 52425 J\"ulich, Germany}
\ead{s.rode@fz-juelich.de, j.elgeti@fz-juelich.de, g.gompper@fz-juelich.de}
\vspace{10pt}

\begin{abstract}
The dynamics and motion of multi-ciliated microswimmers with a spherical body and a small number $N$ (with
$5 < N < 60$) of cilia with length comparable to the body radius, is investigated by
mesoscale hydrodynamics simulations. A metachronal wave is imposed for the cilia beat, for which the 
wave vector has both a longitudinal and a latitudinal component. The dynamics and motion is characterized 
by the swimming velocity, its variation over the beat cycle, the spinning velocity around the main body axis, 
as well as the parameters of the helical trajectory.
Our simulation results show that the microswimmer motion strongly depends on the latitudinal wave
number and the longitudinal phase lag. The microswimmers are found to swim smoothly and usually spin around 
their own axis. Chirality of the metachronal beat pattern generically generates helical trajectories. 
In most cases, the helices are thin and stretched, i.e. the helix radius is about an order of magnitude smaller 
than the pitch. The rotational diffusion of the microswimmer is significantly
smaller than the passive rotational diffusion of the body alone, which indicates that the extended cilia 
contribute strongly to the hydrodynamic radius.
The swimming velocity $v_{swim}$ is found to increase with the cilia number $N$ with a slightly sublinear 
power law, consistent with the behavior expected from the dependence of the transport velocity of planar 
cilia arrays on the cilia separation.
\end{abstract}

%
%
\submitto{\EJP}
%

%
\ioptwocol

\section{Introduction}

Cilia and flagella are the ubiquitous machinery in eukaryotic cells and organisms to generate fluid flow and to propel 
cells and microorganisms in a fluid environment \cite{elge15b, wan21}. While flagella have the beat pattern of a 
sinusoidal traveling wave, and are usually employed to propel single cells like sperm \cite{alva14}, cilia have two 
distinct phases in their beat cycle -- the power and the recovery stroke --, and often work together in pairs like 
in Chlamydomonas reinhardtii \cite{poli:09}, or in large cilia carpets. 
Examples for the concerted action of cilia in carpets are the transport of mucus in the airways \cite{slei88, lois20}, 
the flow generation of the cerebrospinal fluid in brain ventricles \cite{guir10, faub16, pell20}, and the swimming of multi-ciliated 
microorganisms, such as the green alga {\em Volvox} \cite{dres09}, the protozoan {\em Paramecium} \cite{mach72}, and the
placidozoan {\em Opalina} \cite{tamm70}.

A remarkable feature of cilia carpets is that the beat is highly coordinated in the form of metachronal waves
\cite{knig:54, tamm70, guir07, mich10, Osterman2011, Elgeti2013}, where the beat cycles of neighboring cilia have a fixed phase shift.
The power-stroke direction can be parallel or antiparallel to the propagation direction of the metachronal wave, which is denoted
as symplectic or antiplectic wave, but can also point right-wise or left-wise from the wave direction, which is denoted
dexioplectic or laeoplectic wave. The later wave form clearly requires some chirality in the system, which can either be
in the aplanarity of the ciliary beat \cite{Eloy2012}, or result from the spatial arrangement of cilia.

The origin and effect on the transport efficiency of the ciliary beat and of the metachronal wave is fascinating, and has thus
been investigated intensively. It is now well established that hydrodynamic interactions are strong enough 
to cause coordination between neighboring cilia \cite{brum:14, wan14, maes18, hami19, man20}, and suffice to 
explain the formation of metachronal waves in cilia arrays \cite{Elgeti2013}. 
It has also been shown that the transport efficiency can be much higher than for perfect beat synchronization, related to 
the fact that always a fraction of the cilia is in the power stroke, thus avoiding a forward-backward motion of the fluid. 
Additionally, flow generated by the power stroke at the wave crest experiences only small resistance from cilia in the 
neighbouring wave trough, which can all remain close to the anchoring surface during the recovery stroke -- both without 
much steric hindrance \cite{guir07, Eloy2012, Elgeti2013}. 
A further interesting issue is the coordination of the beat directions in cilia carpets, which is now believed to be a 
self-organized process mediated by the fluid flow \cite{guir07, guir10, hami19, man20, lois20}. On the other hand, 
the synchronization of the beat of the two flagella of Chlamydomonas reinhardtii arise from an elastic mechanical 
coupling at their basal foot \cite{wan_coordinated_2016, guo_intracellular_2021}.

More complex is the cilia coordination in multi-ciliated spherical or spheroidal microorganisms. One reason is the well-know
``hairy-ball" theorem, which states that there is no non-vanishing continuous tangent vector field on a surface of spherical
topology \cite{poin85}. This implies that the power-stroke directions of neighboring cilia cannot be parallel everywhere on
a spherical surface, but there have to be at least two defects, which can either be of hedgehog or of swirl type. A second 
reason is that plane metachronal waves are not possible on curved surfaces.

Volvox is a perfect model system for experimental studies of swimming and cilia synchronization 
\cite{brum:12, brum15, Solari2011}. These studies reveal the existence of a symplectic metachronal wave \cite{brum:12}, 
and that the average metachronal coordination is punctuated by periodic phase defects during which synchrony is partial and 
limited to specific groups of cells \cite{brum15}. Under conditions of decreasing nutrient concentration, Volvox colonies were
found to grow larger and increase their flagellar length, separating the somatic cells further, with
the opposing effects of increasing beating force and flagellar spacing balance not significantly affecting the fluid speed 
at the colony surface \cite{Solari2011}.

The theoretical description of the swimming of multi-ciliated microorganisms has lead to the early development of the squirmer
model \cite{ligh52, blak71b}, in which the effect of the ciliary beating is mimicked by a prescribed surface velocity.
This model has been generalized more recently to spheriodal shapes \cite{thee16}, and is employed nowadays to describe the
collective swimming behavior of many types of microswimmers \cite{elge15b, thee18}. The squirmer model has also been
generalized to capture the effect of metachronal waves and to include azimuthal swirl on the continuum level \cite{pedl16b}. 
This model predicts mean swimming speeds and angular velocities as a function of the colony radius qualitatively correct, 
but underestimates both velocities quantitatively \cite{pedl16b}.

The squirmer model applies in the limit that the cilia length and the separation of their anchoring points on the surface
is much smaller than the body size.
The dynamics and flow generation of individual cilia becomes more important in the opposite limit. A model with explicit 
cilia and a prescribed metachronal wave has been introduced recently \cite{ito19}. The model facilitates the calculation 
of hydrodynamic interactions between cilia and the cell body under free-swimming conditions. An antiplectic metachronal 
wave is predicted to be optimal in the swimming speed with various cell-body aspect ratios, which is consistent with 
former theoretical studies \cite{blak72, mich10}. 
The swimming velocity of model ciliates was well represented by the squirmer model. 
The effect of oblique wave propagation is also briefly touched, and is found to lead to a helical swimming 
trajectory \cite{ito19}. 

\begin{figure*}
	\centering
	\subfloat[2-3-2 swimmer]{\includegraphics[width=0.32\textwidth]{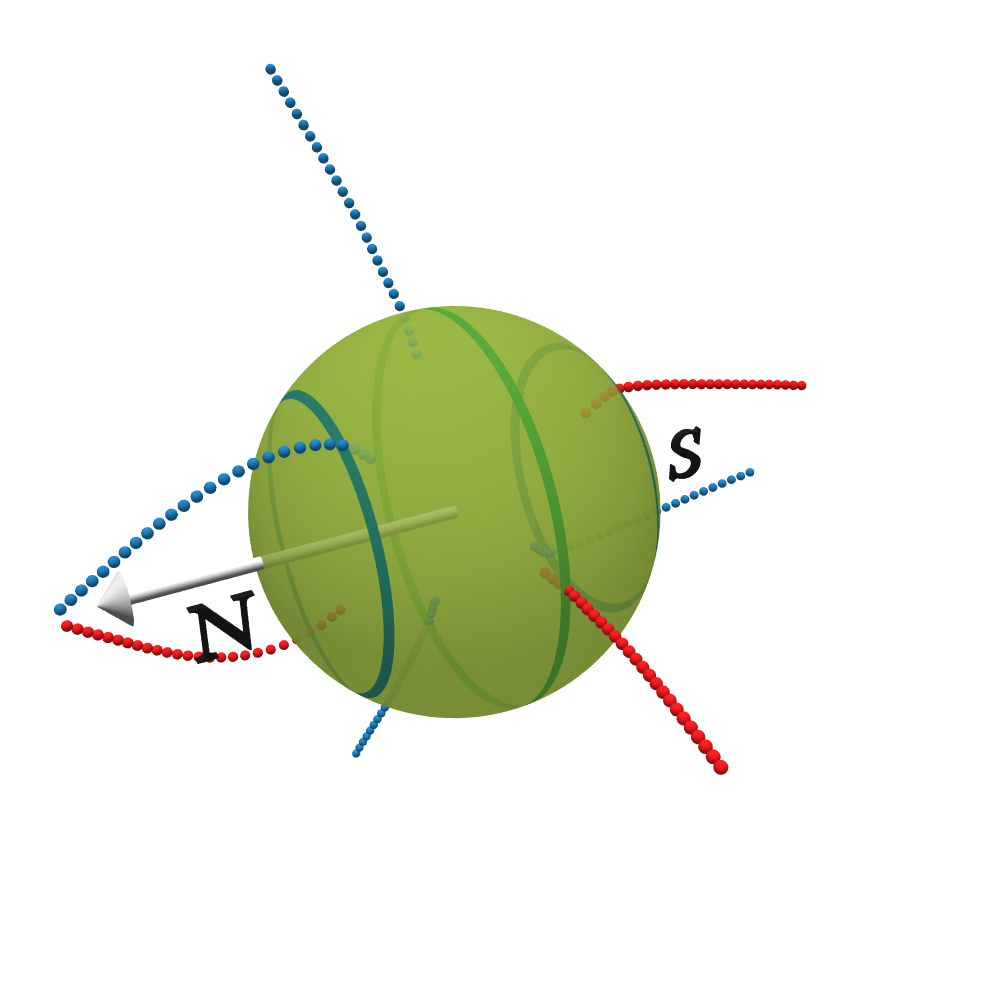} }  
	\subfloat[5-7-5 swimmer]{\includegraphics[width=0.32\textwidth]{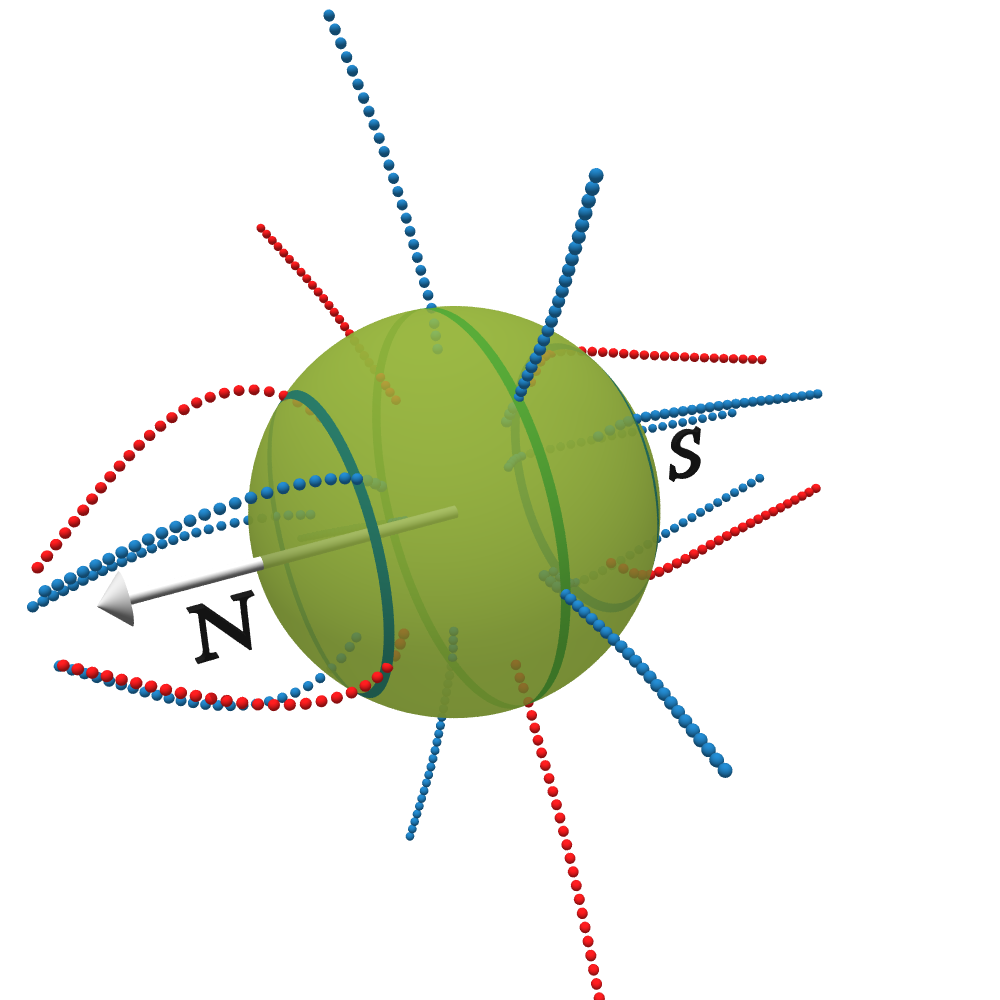} }  
	\subfloat[7-10-7 swimmer]{\includegraphics[width=0.32\textwidth]{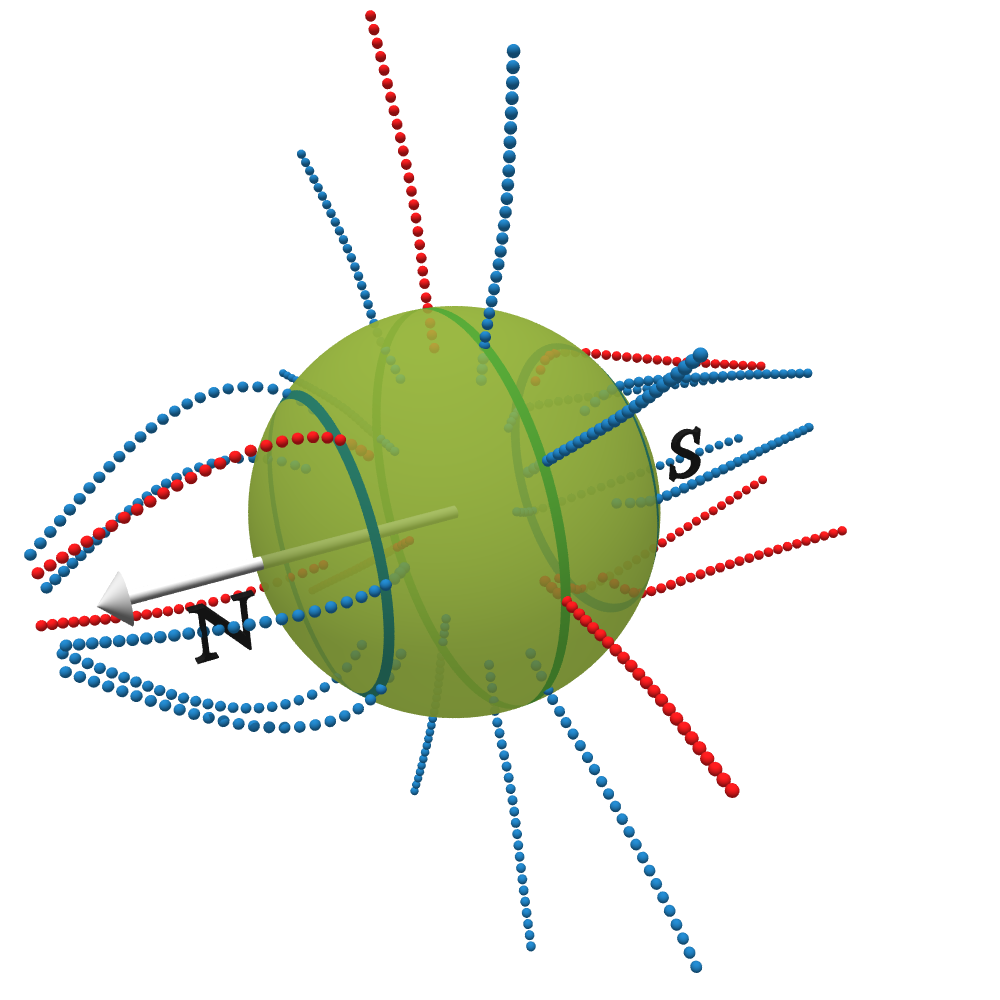} }  
	\caption{Three swimmers with an antiplectic metachronal wave of phase lag $\chi=77^\circ$ with different numbers of 
		cilia placed at the equator, with $\theta=0.0$, and two rings at $\theta=\pm \pi/4$.}
	\label{fig:3swimmer}
\end{figure*}


Also, there has been significant progress recently on the experimental techniques for cilia characterization \cite{cicu20},
as well as the construction of artificial, externally actuated cilia carpets \cite{gu20}, which relates to the 
goal of the construction of soft microbots \cite{Gompper_Roadmap_2020}. Soft robots with antiplectic waves have been
shown to exhibit much higher locomotion speed than those with symplectic waves \cite{gu20}.

We employ a similar model of ciliated microorganisms as studied in Ref.~\cite{ito19}, but focus on the regime of a 
smaller number of longer cilia -- comparable in length to the body size. Volvocalean algae, which falls into this 
parameter range are, for example, the 16-celled {\em Pandorina} and the 32-celled Eudorina \cite{kirk05, Solari2011}.
We focus on the efficiency of metachronal waves for translational and rotational motion of the swimmer.  In particular, 
we investigate the influence of the wave direction, the phase lag between neighboring cilia, and the cilia density, 
on the swimming efficiency and the persistence of the swimming trajectory. In particular, we identify parameter 
combinations for which helical swimming trajectories emerge.


\section{Ciliated microswimmer and hydrodynamic simulation}
\label{sec:model}

The ciliated microswimmer is modeled as a spherical body of radius $R$, to which several cilia are attached 
(see Fig.~\ref{fig:3swimmer}). 
The body consists of $N_b$ point particles covering the surface of a sphere, with an additional particle at its 
center. The particles on the surface are connected to their neighbors by stiff harmonic springs
to form a triangular network, as well as to the center particle to form an essentially rigid sphere. 
Following Refs.~\cite{elge10, Elgeti2013}, we model a cilium of length $L$ by three semi-flexible polymers, which 
are arranged on the surface of a (hypothetical) cylinder with triangular cross section. Each semi-flexible
polymer is described by a bead-spring model, consisting of $N_c$ beads connected by harmonic springs with rest 
length $\ell_c$ and spring constant $k_c$. 
The three polymers are interconnected by springs in order to retain the cylindrical shape over time. 
The cilia are anchored in the body with a ``clamped" boundary condition, which is achieved by
extending the filament by a segment of length $R/4$ inside the body. This anchoring is implemented by
stiff springs which connect the sub-surface part of the cilium to the sphere surface, mimicking the embedding 
of the basal bodies of the cilia in the cell body. In detail, 
the first and fourth particle of each of the three polymers constituting  the cilium are connected to the 
closest particle on the sphere surface as well as its next-nearest neighbors. This construction anchors the 
cilium tightly to the body without a noticeable body deformation -- even for the largest applied ciliary forces.

\begin{figure}[t]
	\centering
	\includegraphics[width=0.98\columnwidth]{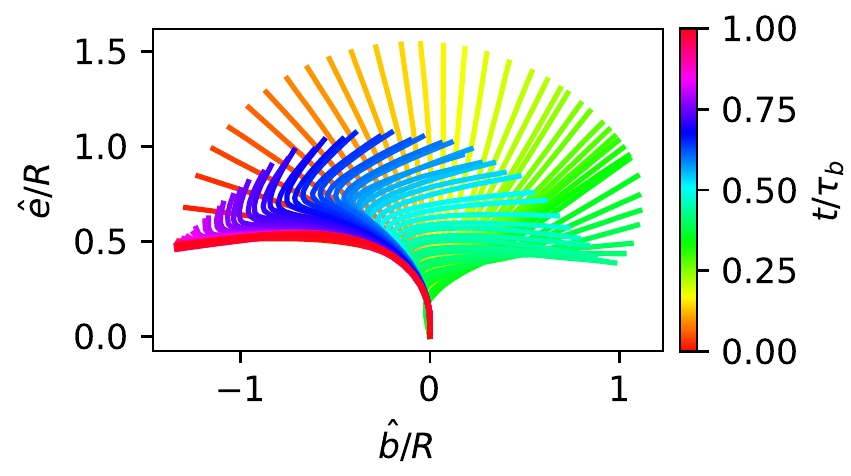}
	\caption{Sketch of the ciliary beat pattern - The color indicates time. The elongated conformation during the power stroke 
		(orange to green) is followed by the buckled conformation during the recovery stroke (green to blue). }
	\label{fig:beat_pattern}
\end{figure}

The cilia beat pattern consists of a power and a recovery stroke (see Fig.~\ref{fig:beat_pattern}). 
The ciliary beat is generated by varying 
the equilibrium spring lengths $\ell_c$ of one selected polymer, both spatially along the cilium and periodically 
in time, which creates a spatially and temporarily varying cilium curvature. The selection of the active polymer 
defines the beat plane. 
The beat pattern of power and recovery stoke is obtained by prescribing an analytic function for the desired local
curvature of the cilium. For details, see \ref{app:cil_beat}. This is inspired  by the molecular mechanism 
which drives ciliary beating, where molecular motors apply torques along the flagellum, but each motor 
has a maximum force it can generate. 
To mimic the stall force, and to avoid artificial cilia shapes due to unnaturally large local torques, we limit 
the change of equilibrium bond length such that a maximum energy of $1.0~k_B T$ per MPC time step (see below) 
can be inserted into the system, where $k_BT$ is the thermal energy.
The dynamics of the beat is not only determined by the time-dependence of the internal torques, but is also 
affected by the flow field around the swimmer and the elastic properties of the cilium. To model the hydrodynamics of
the embedding fluid, we employ multi-particle collision dynamics (MPC), a mesoscale simulation technique,
which is ideally suited for simulations with a particle-based model of an active microswimmer. In this approach,
the fluid consists of point particles, each characterized by its location ${\bf r}_i(t)$ and velocity ${\bf v}_i(t)$.
These particles move ballistically during the streaming step for a time interval $h$. In the subsequent
collision step, all particles are sorted into the cells of a simple cubic lattice with lattice constant $a$. 
Particles in each cell interact by exchanging momentum, but in such a way to conserve the mass and linear momentum 
within each cell. A cell-level canonical thermostat (with Maxwell-Boltzmann scaling) is applied after every 
collision step to maintain a constant temperature $T$ \cite{Huang_PRE_2015}. 
A detailed description of the MPC technique and a review of its application to many systems in soft, active and 
living matter is provided in Refs.~\cite{gomp09, Winkler_MotActMat_2018, Shaebani2020}. 

For the investigation of the self-organization of beating cilia arrays into metachronal waves in 
Ref.~\cite{Elgeti2013}, the switch from power to recovery stroke, and back, needs triggers, 
which are reached faster or slower depending on the environmental conditions (``geometrical clutch" hypothesis 
\cite{lind14}). In Ref.~\cite{Elgeti2013}, a critical value curvature of 
the cilium, and a maximum angle between the extended cilium and the normal vector to the cell surface, were 
employed to trigger these switching events. 
We use here a variant of this model, in which the force generation  is modified
such that the switch between power and recovery stroke becomes deterministic in time, with a power-stroke 
time $\tau_p$ and a recovery-stroke time $\tau_r$, with $\tau_r > \tau_b$, which results in a constant beat 
period $\tau_b = \tau_p + \tau_r$. In this case, a phase lag between the beats of neighboring cilia is imposed
to induce a metachronal wave (see below).

\begin{figure*}[t]
	\centering
	\includegraphics[width=0.90\textwidth]{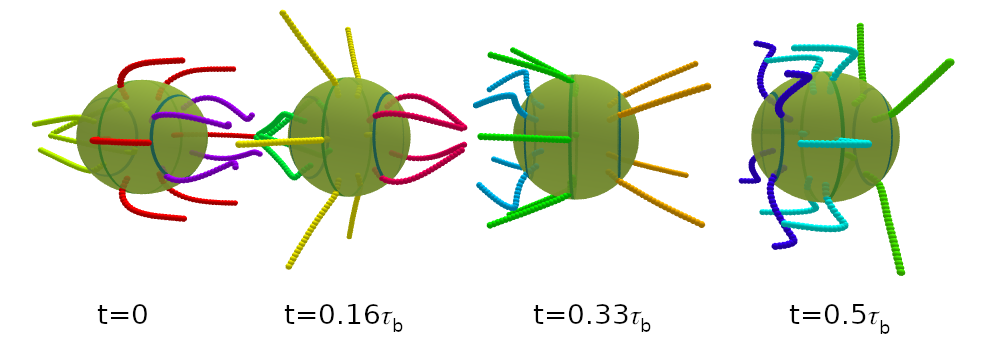} 
	\caption{Temporal beat pattern of a 4-6-4 swimmer with cilia beat in the longitudinal direction 
		($\theta_r=0$), with $k_\varphi=0$ and phase lag $\chi=-77^\circ$. The varying cilium color indicates 
		the instantaneous stage in the beat cycle (compare Fig.~\ref{fig:beat_pattern}). 
		The progression of the beat is indicated by the time $t$, given in units of the beat period $\tau_b$. 
		The motion of the swimmer is from left to right. The translational motion is not to scale. 
		See also movie SM1 for an illustration of the swimming behavior. 
	}
	\label{fig:movie-stills-thetar0_k0}
\end{figure*}

\begin{figure*}[t]
	\centering
	\includegraphics[width=0.90\textwidth]{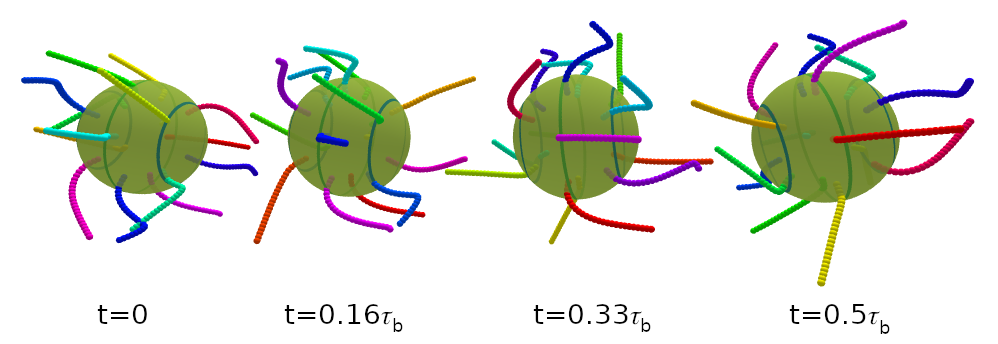} 
	\caption{Temporal beat pattern of a 4-6-4 swimmer with cilia beat in the longitudinal direction 
		($\theta_r=0$), with $k_\varphi=1$ and phase lag $\chi=-77^\circ$. The varying cilium color indicates 
		the instantaneous stage in the beat cycle (compare Fig.~\ref{fig:beat_pattern}). 
		The progression of the beat is indicated by the time $t$, given in units of the beat period $\tau_b$. 
		The motion of the swimmer is from left to right. The translational motion is not to scale. 
		See also movie SM2 for an illustration of the swimming behavior. 
	}
	\label{fig:movie-stills-thetar0_k1}
\end{figure*}

We consider a ciliated microswimmer, where a total of $N$ cilia are placed equidistantly on three latitudinal rings. 
In spherical coordinates $(\varphi, \theta)$, one ring is located at the equator, $\theta=0$, and two rings 
at $\theta=\pm \pi/4$, see Fig.~\ref{fig:3swimmer}. In order to have a roughly constant cilia density, the 
number of cilia at the ``polar circles" is reduced by a factor close to $1/\cos \theta = \sqrt{2}$ compared to their
number $N_{eq}$ at the equator. The ``first"
cilium on each ring is initially placed at $\varphi=0$; subsequently, the cilia positions on both polar rings 
are shifted azimuthally in the same direction by a small $\Delta \varphi= \pi/N_{eq}$ to avoid 
perfect registry for one particular longitude. Several examples of microswimmer with various numbers of cilia 
are shown in Fig.~\ref{fig:3swimmer}). 

The power-stroke direction of the cilia can be along ${\bf e}_\theta$, i.e. be parallel to the circles of longitude,
or deviate from this highly symmetric case. This is modeled by rotating the beat plane of each cilium around the 
radial axis ${\bf e}_r$ by an angle $\theta_r$ (with $\theta_r=0$ corresponding to the longitudinal beat direction).
A beat-plane orientation with $\theta_r \ne 0$ introduces chirality in the swimmer propulsion pattern. 

In the case of an imposed metachronal wave, we define a local phase $\Psi(\varphi, \theta)$ for each cilium on the surface 
of the sphere,
\begin{equation}
\Psi(\varphi, \theta) = k_\varphi \varphi + k_\theta \theta,
\end{equation}
with $-\pi < \theta < \pi$ and $0\le \varphi <2\pi$, 
where the direction of the wave is determined by the wave vector ${\bf k}=(k_\varphi, k_\theta)$, which has 
longitudinal and latitudinal components, $k_\theta$ and $k_\varphi$.
Since we want to have continuous wave solution traveling around the circles of constant latitude, $k_\varphi$ has to be an 
integer number.
For $N_{eq}$ equally-spaced cilia on the equator, $k_\varphi=N_{eq}/2$ results in a phase lag between neighboring cilia 
of $\Delta \Psi = \pi$. Since $k_\varphi \in \mathbb{N}$, we limit it to $k_\varphi=0,1,...,(N_{eq}//2+1)$, 
where $//$ denotes integer division. 
The longitudinal wave vector $k_\theta \in \mathbb{R}$ is not restricted by any physical boundary conditions. We employ 
the phase lag 
\begin{eqnarray}
	\chi = k_\theta\, \Delta \theta = k_\theta \frac{\pi}{4},
\end{eqnarray}
between neighboring rings to quantify $k_\theta$,
where $\Delta\theta=\pi/4$ is the fixed latitudinal angular distance between successive rings. 

For zero longitudinal wave component, $k_\varphi=0$, a positive phase lag $\chi$ corresponds to a symplectic metachronal wave, 
which travels in the direction of the power stroke, a negative phase lag $\chi$ to an antiplectic metachronal wave, 
which travels in the direction of the recovery stroke. 
For non-zero longitudinal component $k_\varphi$, the metachronal wave becomes either dexioplectic  ($k_\theta > 0$) 
or laeoplectic ($k_\theta < 0$). Because the microswimmer is constructed essentially symmetric to the main axis, dexioplectic 
waves show the same effect on the propulsion as laeoplectic, except for an opposite direction of axial rotation $\Omega_n$.
Therefore, we restrict our analysis to symplectic, antiplectic ($k_\varphi=0$) and dexioplectic ($k_\varphi > 0$) 
metachronal waves with varying phase lags $\chi$.

For simulations, we employ the following parameters. The spherical body consists of $N_b=643$ mesh points,
the cilia are constructed with $N_c=26$ beads for each polymer strand, they have length (outside the body) of $L=3R$, 
and the body radius in terms of MPC collision box size $a$ is $R=8a$, which guarantees a good resolution of the
hydrodynamic flow fields. The MPC simulations employ collisions described by stochastic rotation dynamics, 
with a time step $h=0.05$, cell size $a=1$, rotation angle $\alpha_0= 130^\circ$, and an overall simulation box size 
of $(100a)^3$.

\section{Results}
\label{sec:results}

\subsection{The 5-7-5 Swimmer with Longitudinal Beat Direction}
\label{sec:longitudinal}

We focus on the swimming properties of a spherical swimmer with 7 cilia on the equator and 5 cilia on the two polar rings 
see Fig.~\ref{fig:3swimmer}, with power stroke direction along the main body axis, $\theta_r=0$.
Examples for the beating dynamics with phase lag $\chi=-77^\circ$ with latitudinal wave numbers $k_\varphi=0$ and $k_\varphi=1$ 
are shown in Fig.~\ref{fig:movie-stills-thetar0_k0} and Fig.~\ref{fig:movie-stills-thetar0_k1}, respectively.

\begin{figure*}
	\includegraphics{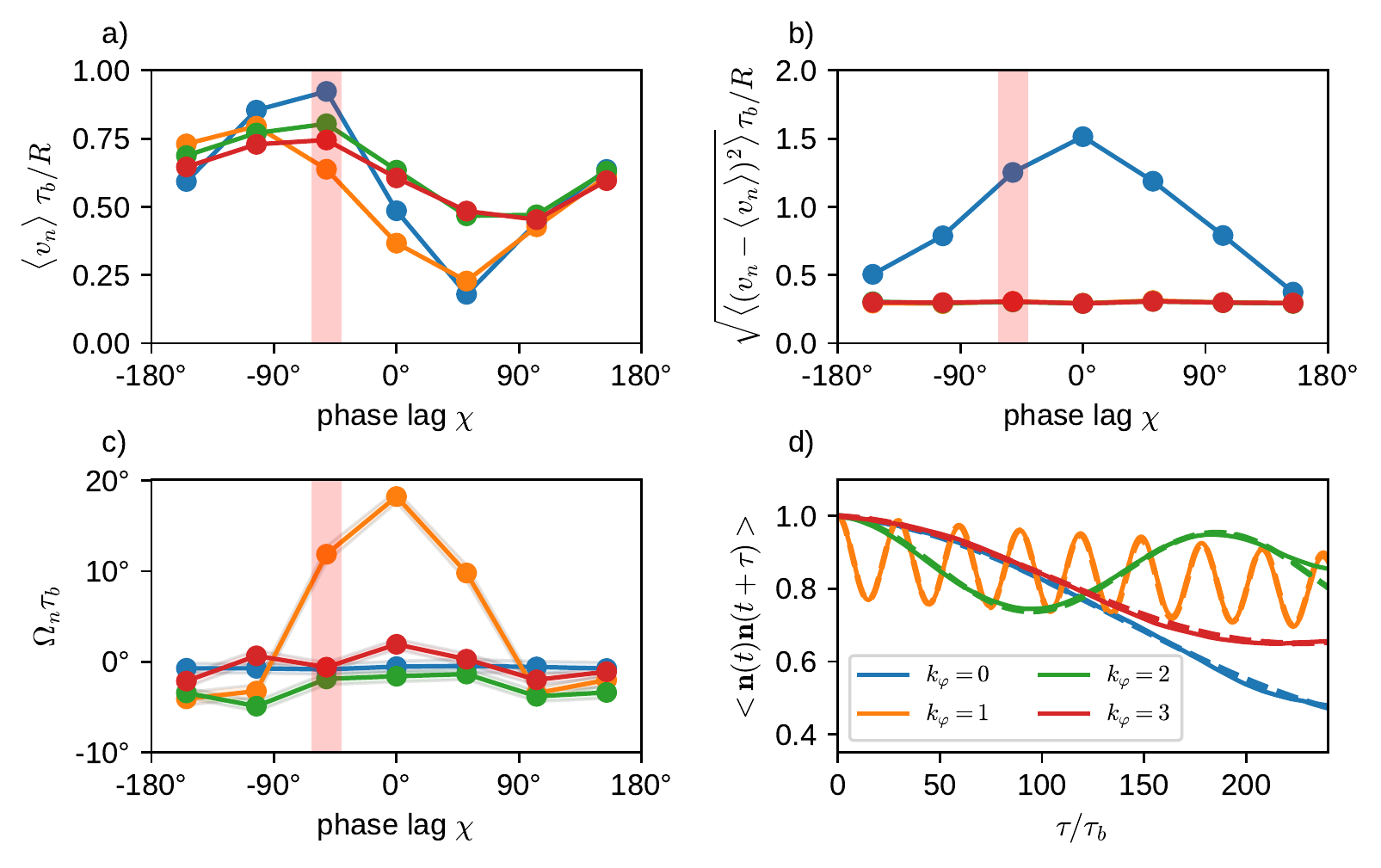} 
	\caption{Swimming properties of  a 5-7-5 swimmer with cilia beat in the longitudinal direction ($\theta_r=0$), 
		for various phase lags $\chi$ and latitudinal modes $k_\varphi$. 
		\textbf{a)} Average swimming velocity $<v_n>$. 
		\textbf{b)} Velocity fluctuations $\sqrt{<(v_n - <v_n>)^2>}$ around the average.  
		\textbf{c)} Rotational velocity $\Omega_n$ around the main body axis. 
		\textbf{d)} Temporal auto-correlation function $<{\bf n}(t)\cdot{\bf n}(t+\tau)>_t$ for swimmers with phase lag 
		$\chi=-50^\circ$ (red stripe). The dashed lines are fits to the auto-correlation function, Eq.~(\ref{eq:correlations}) 
		(see text). Note that in (b), the curves for $k_\varphi=1$ and $k_\varphi=2$ are numerically identical to the
		curve for $k_\varphi=3$, and are therefore not visible.
	}
	\label{fig:vel_n474_parallel}
\end{figure*}

We consider the propulsion velocity $\langle v_n \rangle$, the velocity fluctuations 
$\sqrt{\langle (v_n - \langle v_n \rangle)^2 \rangle}$ around the average,
and the rotational velocity $\Omega_n$ around the main body axis, see Fig.~\ref{fig:vel_n474_parallel}.
All these quantities depend on both the phase lag $\chi$ in the longitudinal direction and the wave number $k_\varphi$ 
in the latitudinal direction.
The propulsion velocity $\langle v_n \rangle$ in the direction of the main body axis shows a pronounced ``sinusoidal" dependence 
on the phase lag $\chi$ for all wave numbers $k_\varphi$, see Fig.~\ref{fig:vel_n474_parallel}a. Here, the strongest variation 
is found for a metachronal wave with $k_\varphi=0$. 
The swimming velocity $v_n$ is maximal for a negative phase lag of $\chi \simeq -50^\circ$ 
(see Fig.~\ref{fig:vel_n474_parallel}a) and reaches almost a body length per beat cycle, $v_n = 0.9 R/\tau_b$. 
Such a negative phase lag corresponds to an antiplectic metachronal wave, with cilia of one ring lagging behind those of
the subsequent ring by a little bit more than 1/4 of a beat cycle, where the wave travels against the direction of the power 
stroke. The smallest velocity of $v_n = 0.2 R/\tau_b$ corresponds to symplectic wave with a positive 
phase lag of $\chi\simeq50^\circ$, where the wave travels with the direction of the power stroke.
These results are consistent with former theoretical studies of efficiency optimization \cite{mich10, Osterman2011}.

The variation $\sqrt{\langle(v_n - \langle v_n \rangle)^2 \rangle}$ of the swimming velocity is shown 
in Fig.~\ref{fig:vel_n474_parallel}b.
The swimming velocity $v_n$ fluctuates mostly for swimmers with $k_\varphi=0$, whereas it is nearly independent of the 
phase lag $\chi$ for latitudinal wave numbers with $k_\varphi>0$. In particular for the synchronous case, with $\chi=0$, 
the swimmer moves quickly forward during the power stroke, but reverses its direction of motion during the recovery stroke, 
which in sum leads to a relative slow average velocity with high fluctuations.

For $k_\varphi \ge 1$, the latitudinal component of the wave leads to an additional rotation velocity $\Omega_n$ around 
its main body axis (see Fig.~\ref{fig:vel_n474_parallel}c). 
The rotation is most pronounced for $k_\varphi=1$ and $-50^\circ < \chi < 50^\circ$, and is very small for 
$k_\varphi \ge 2$ and all phase lags $\chi$. This behavior can be understood by considering two contributions.
(i) For $k_\varphi \ge 1$, rotation is enhanced when the metachronal wave travels in the latitudinal direction, 
i.e. $\chi\simeq 0$. (ii) Latitudinal wave numbers $k_\varphi \ge 2$ imply short metachronal 
wave lengths, which cannot propel fluid effectively, because the opposing beat of neighboring
cilia just generates local swirls. For example, $k_\varphi=3$ corresponds to a phase lag between neighboring cilia 
on the equator of $\Delta\Psi= k_\varphi 2\pi/N_{eq}$, which implies a large phase lag of 
$\Delta\Psi\simeq 150^\circ$ (for $N_{eq}=7$). 
The rotational component of the propulsion also slightly reduces the component that contributes to the swimming 
velocity, compare Fig.~\ref{fig:vel_n474_parallel}a.

\begin{figure*}
	\centering
	\includegraphics[width=0.48\textwidth]{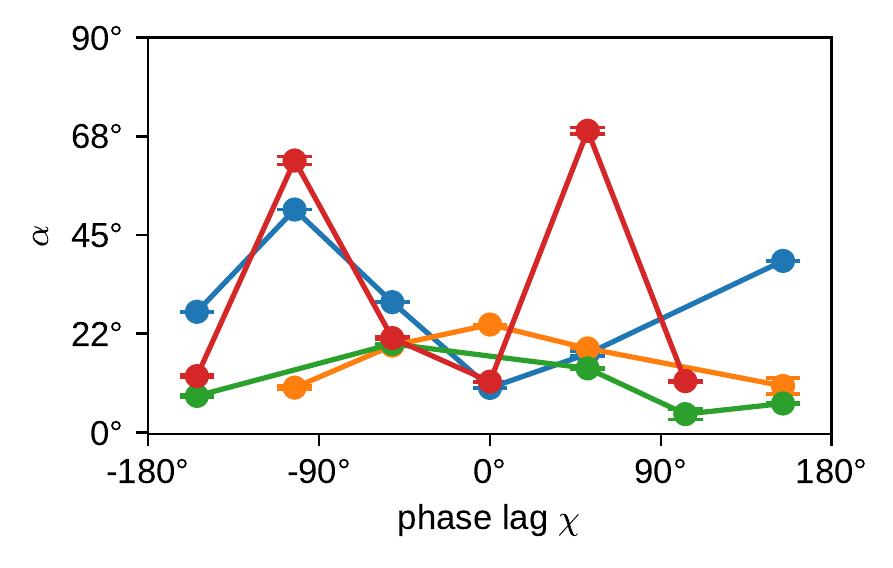} 
	\includegraphics[width=0.48\textwidth]{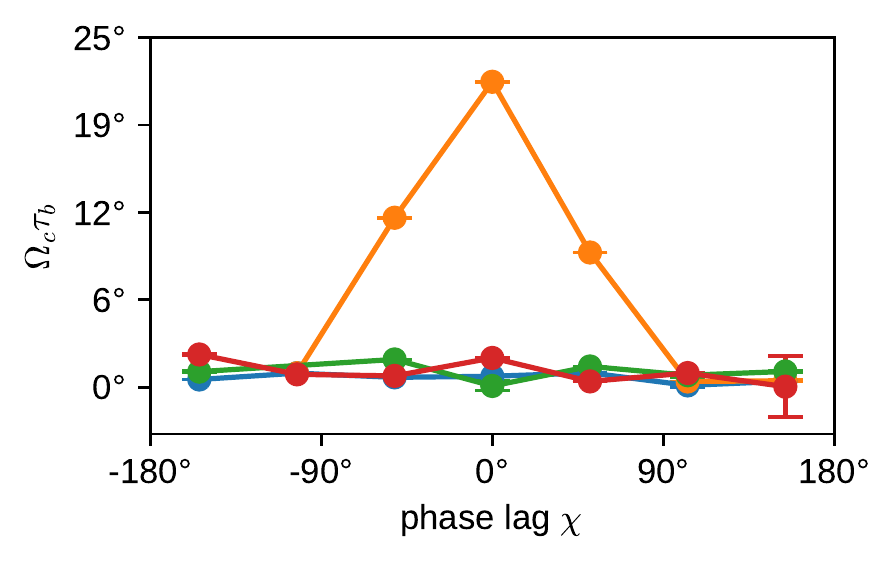} 
	\includegraphics[width=0.48\textwidth]{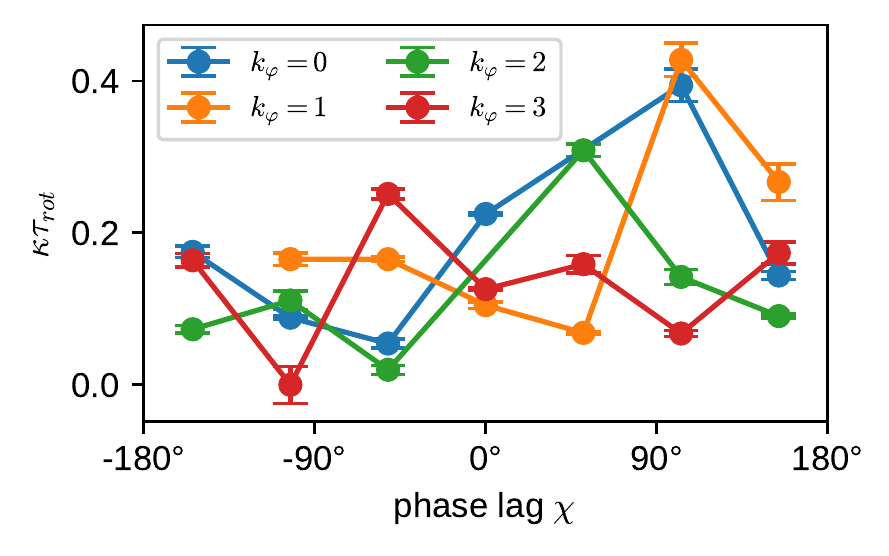} 
	\includegraphics[width=0.48\textwidth]{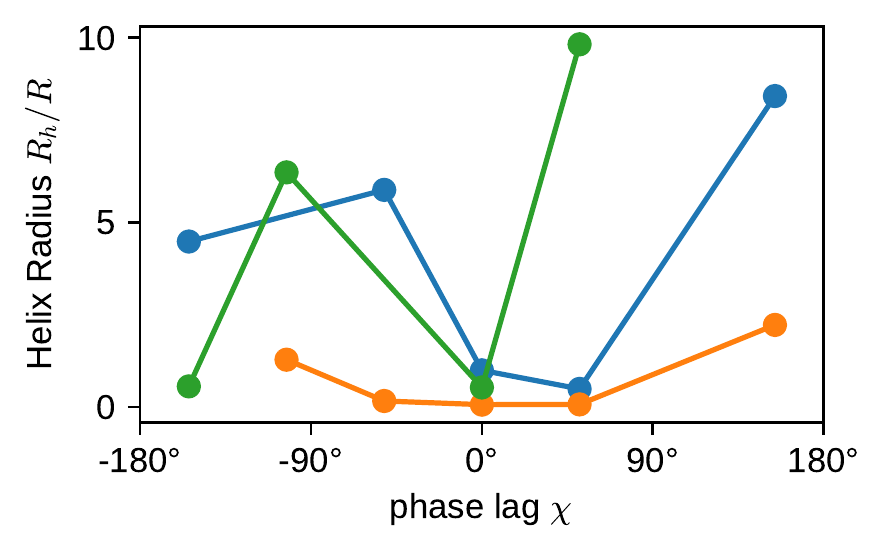} 
	
	\caption{Swimming properties of  a 5-7-5 swimmer with cilia beat in the longitudinal direction ($\theta_r=0$), 
		for various phase lags $\chi$ and latitudinal modes $k_\varphi$. 
		\textbf{a)} Alignment angle $\alpha$;
		\textbf{b)} Rotation frequency $\Omega_c$, normalized by the beat period $\tau$;
		\textbf{c)} Inverse correlation time $\kappa$, normalized by the rotational diffusion time $\tau_{rot} = 1/(2D_{rot})$.
		\textbf{d)} Helix radius $R_h$ in units of the body radius $R$, as obtained from Eq.~(\ref{eq:helix_params}).
	}
	\label{fig:correlation_parameters_theta_r0}
\end{figure*}

For $k_\varphi \ge 1$, the chirality of the wave pattern not only induces a body rotation, but also implies a helical swimming 
trajectory \cite{cren89,elge10}.
Consider a swimmer, for which the main body axis ${\bf n}$ rotates around a fixed axis ${\bf e}_\parallel$ in the lab 
reference frame with an opening angle $\alpha$. In the absence of translational and rotational noise, this corresponds to the
trajectory of a perfect helix with axis ${\bf e}_\parallel$ and azimuthal direction ${\bf e}_{\perp}(t)$
\begin{eqnarray}
{\bf n}(t) &=& n_{\parallel} {\bf e}_\parallel +  n_\perp {\bf e}_{\perp}(t) \cr
 n_\parallel &=& \cos \alpha \cr
 n_\perp &=& \sin \alpha 
 \label{eq:helix}
\end{eqnarray}
where ${\bf e}_{\perp}(t)=(\cos(\Omega_c t), \sin(\Omega_c t))$ in Cartesian coordinates in the plane with normal vector 
${\bf e}_\parallel$.

The opening angle $\alpha$, the rotation frequency $\Omega_c$, and swim velocity $v_n$, are related to the helix parameters -- 
helix radius $R_h$, pitch length $P_h$, an helix angle $\alpha_h$ -- as
\begin{eqnarray}
	\cos(\alpha)^2 = P_h^2/[P_h^2 + 4\pi^2 R_h^2], \cr 
	\alpha_h = \arctan(2\pi R_h/P_h) \equiv \alpha,
	\label{eq:helix_geom}
\end{eqnarray}  
and
\begin{equation}
	P_h = v_n \cos(\alpha)/\Omega_c, \ \ \  R_h = v_n \sin(\alpha)/(2\pi \Omega_c).
\label{eq:helix_params}
\end{equation}
For $\alpha=0$ the microswimmer moves on a straight lines, whereas for $\alpha=90^\circ$ it moves on a circle. 
In general, the directional auto-correlation function of a swimmer consists of two factors, the correlation due to the 
helical motion, which depends on the inclination angle $\alpha$, and an exponential decay due to thermal or active noise, 
\begin{equation}
<{\bf n}(t) \cdot {\bf n}(t+\tau)> = \left(\cos^2 \alpha + \sin^2 \alpha \cos(\Omega_c \tau)\right ) e^{-\kappa \tau}
\label{eq:correlations}
\end{equation}

Correlation functions for fixed $\chi=-50^\circ$ and various $k_\varphi$ are shown in Fig.~\ref{fig:vel_n474_parallel}d.
Results for the fitted parameters of the helical motion and the decay time $1/\kappa$ in Eq.~(\ref{eq:correlations})
are displayed in Fig.~\ref{fig:correlation_parameters_theta_r0}. The calculation of these parameters requires very
long simulation times of several hundred beat periods $\tau_b$ or more. Even in this case, trajectories are 
sometimes too short for a reliable parameter estimation, in particular for large decorrelation times $1/\kappa$.

\begin{figure}
	\includegraphics[width=0.48\textwidth]{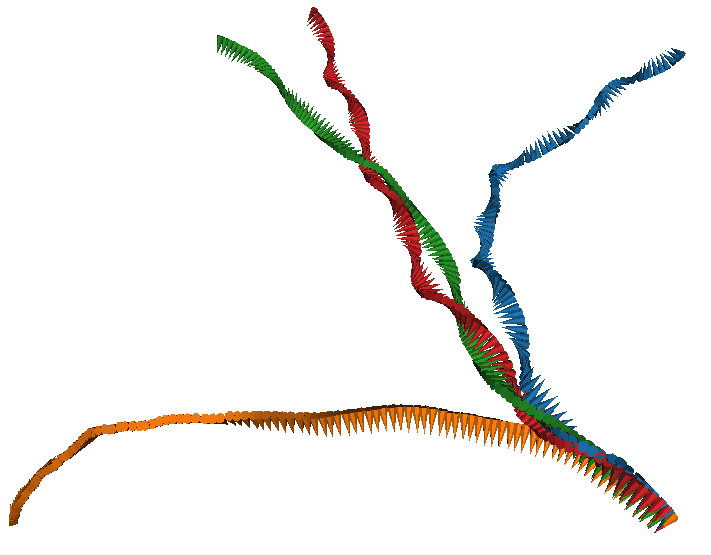} 
	\caption{Helical trajectories for selected parameters of the metachronal beat, with phase lag $\chi=+105^\circ$.
		Trajectory colors indicate the latitudinal wave number, with  $k_\varphi=0$ (blue), $k_\varphi=1$ (orange), 
		$k_\varphi=2$ (green), and $k_\varphi=3$ (red). The small (body-fixed) arrows have their tip in the body center and 
	    their base on the equator to indicate rotation around the main body axis.}
	\label{fig:trajectories}
\end{figure}

For most cases, the helix angle $\alpha < \pi/4$, which means that the constant term in Eq.~(\ref{eq:helix}) 
dominates over the oscillatory term, and thus the helix is thin and elongated. Only in a few cases, like $k_\varphi=3$,
we find a nearly circular motion and a tightly wound helix.
The circling frequency $\Omega_c$ is pronounced for $k_\varphi = 1$, and small phase lags $|\chi|$, which leads to
a pronounced helical swimming trajectory, see Fig.~\ref{fig:trajectories}.
This is closely related to the large internal spinning frequency $\Omega_n$ (see Fig.~\ref{fig:vel_n474_parallel}c).
For $k_\varphi \ge 2$,  the circling frequency is typically very small, which is again related to the inefficiency of
propulsion for short metachronal wave lengths. 
The data in Fig.~\ref{fig:correlation_parameters_theta_r0} sometimes appear to have a larger ``scatter" for different 
wave numbers $k_\varphi$ and different phase lags $\chi$. However, two points should be noticed. (i) We expect smooth curves for 
fixed $k_\varphi$ as a function of $\chi$ when $\chi$ is varied in small steps; however, we vary $\chi$ in rather large 
discrete steps of about $50^\circ$, which results in significantly different metachronal waves. (ii) Similarly,
different $k_\varphi$ generate very different wave patterns, compare Figs.~\ref{fig:movie-stills-thetar0_k0} 
and \ref{fig:movie-stills-thetar0_k1} for $k_\varphi=0$ and $k_\varphi=1$, respectively.

\begin{figure}
	\includegraphics[width=0.48\textwidth]{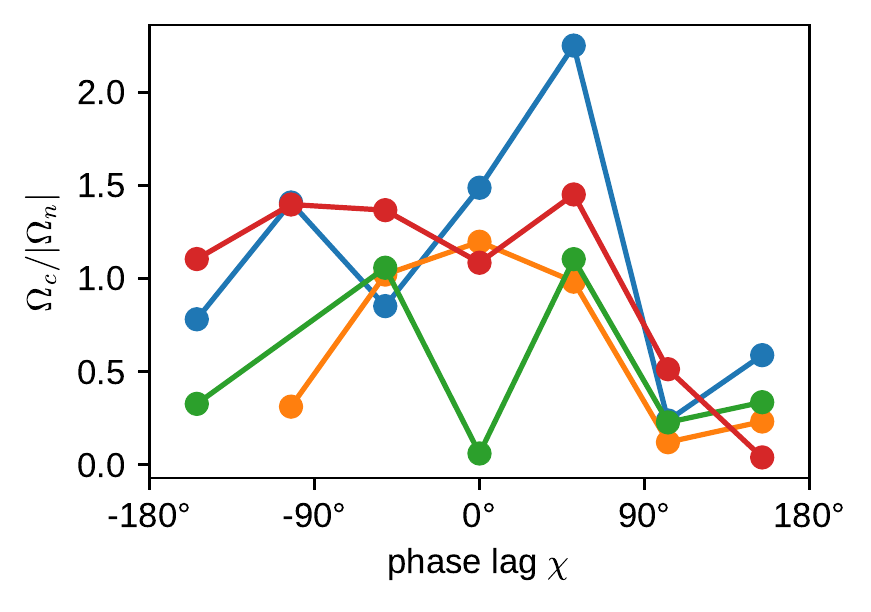} 
	\caption{Ratio of circling frequency $\Omega_c$ and spinning frequency $\Omega_n$ for various latitudinal wave
		numbers $k_\varphi$. The color code is the same as in Fig.~\ref{fig:correlation_parameters_theta_r0}.}
	\label{fig:frequency_ratio}
\end{figure}

The ratio $\Omega_c/|\Omega_n|$ is displayed in Fig.~\ref{fig:frequency_ratio}; it demonstrates that the circling and
spinning frequencies are closely related. In many cases, $\Omega_c/|\Omega_n|\simeq 1$, which corresponds to a 
``twisted-ribbon-like" motion (or to a ``tidal-locking-like" motion, as for the moon, which always presents the
same side to the earth). However, there are also cases where $\Omega_c/|\Omega_n|$ is close to $0$ or to around $2$,
which indicates that the two frequencies don't have to be locked always. An example for $\Omega_c/|\Omega_n|=0$ is
a microswimmer which spins around its body axis, but swims on a straight trajectory.

It is interesting to note that the microswimmers with $k_\varphi=0$ do {\em not} have a vanishing circling frequency.
The reason is that they are not perfectly axisymmetric, and therefore have some inherent chirality, due to the 
non-symmetric location of the cilia on the body. For example, there is a particular longitude, where the cilia in the 
various rings are closest together, which generates a non-axisymmetric flow field. This leads to the helical 
trajectory displayed in Fig.~\ref{fig:trajectories}, with a non-vanishing but very small $\Omega_c$. 
The spinning frequency $\Omega_n$ around the main body axis nearly vanishes in this case (see Fig.~\ref{fig:vel_n474_parallel}c), 
because for $\theta_r=0$ there is essentially no component of the cilia beat in the latitudinal direction. 

Finally, the decay (or decorrelation) time $1/\kappa$ is determined by thermal fluctuations and Stokes friction,
and possibly by active fluctuations generated by the beat. In the thermal case, we expect a rotational 
diffusion time $\tau_{rot} = 1/(2D_{rot})$ with rotational diffusion coefficient $D_{rot}=k_BT/(8\pi \eta R^3)$.
The scaled decorrelation times $1/(\kappa \tau_{rot})$ are mostly larger than $4$, significantly larger than unity.
This implies that (i) activity has a small effect on the rotational diffusion, and (ii) that the cilia contribute
significantly to reduce the rotational diffusion by increasing the effective hydrodynamic radius. A factor $4$ 
reduction is equivalent to a hydrodynamic radius $R_{hydro}= 1.6 R$, which is not implausible as the 
geometric radius from body center to cilia tips is $4R$.
The least persistent motion is observed for $k_\varphi=1$ and $\chi=100^\circ$, which indicates a rather strong 
active contribution to the noise, as might be expected for the strong wiggling motion for $k_\varphi=1$ 
(compare movie SM2). 

We can now use the expressions (\ref{eq:helix_params}) to extract the helix radius $R_h$ 
from the data presented in Figs.~\ref{fig:vel_n474_parallel} and \ref{fig:correlation_parameters_theta_r0}. 
The results are shown in Fig.~\ref{fig:correlation_parameters_theta_r0}(d). They indicate that the helix radius $R_h$ 
is typically rather small, from essentially zero to just a few times the body radius $R$, with a few exceptions, 
like $k_\varphi=0$ and $\chi=150^\circ$. The helix pitch is usually a factor $10$ larger than the radius, as 
typically $P_h/(2\pi R_h) = \cot(\alpha) \gg 1$
(compare Eq.~(\ref{eq:helix_geom}) and Fig.~\ref{fig:correlation_parameters_theta_r0}(a)).
This is in good agreement with the trajectories displayed in Fig.~\ref{fig:trajectories}.

\subsection{The 5-7-5 Swimmer with Oblique Power-Stroke Direction}
\label{sec:oblique}

\begin{figure*}
	\centering
	\includegraphics[width=0.90\textwidth]{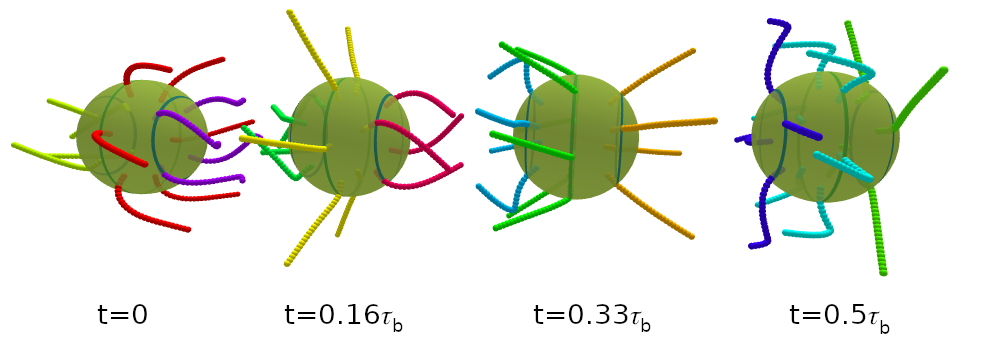} 
	\caption{Temporal beat pattern of a 4-6-4 swimmer with cilia beat in an oblique direction 
             ($\theta_r=22.5^\circ$), with latitudinal wave vector $k_\varphi=0$ and phase lag $\chi=-77^\circ$. 
             The varying cilium color indicates the instantaneous stage in the beat cycle (compare Fig.~\ref{fig:beat_pattern}). 
             The progression of the beat is indicated by the time $t$, given in units of the beat period $\tau$. 
             The motion of the swimmer is from left to right. The translational motion is not to scale. 
             See also movie SM3 for an illustration of the swimming behavior.
	}
	\label{fig:movie-stills-thetar22_k0} 
\end{figure*}

Except for the oblique propagation direction of the metachronal wave, there are other possibilities to achieve a 
chirality of the dynamic beat pattern of a ciliated microswimmer. One of these possibilities is to vary the 
power stroke direction away from the main body axis, and rotate it in the local tangent plane to the left by a
tilt angle $\theta_r$. For simplicity, we consider in this case only a metachronal wave in the main body direction,
i.e. $k_\varphi=0$. An example for the beating dynamics with phase lag $\chi=-77^\circ$ and tilt angle 
$\theta_r=22.5^\circ$ is shown in Fig.~\ref{fig:movie-stills-thetar22_k0}.

\begin{figure*}
	\includegraphics{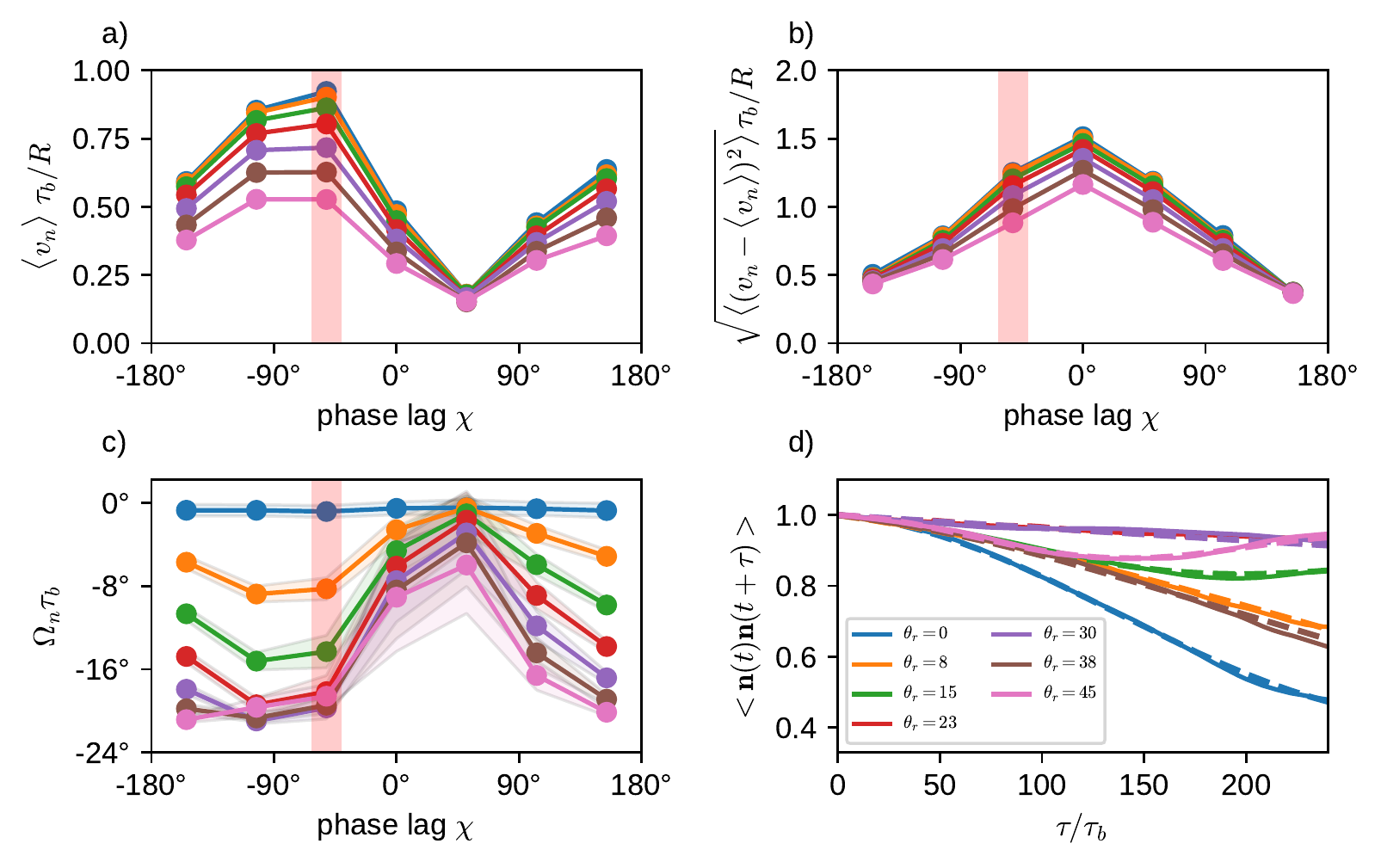} 
	\caption{Swimming properties of a 5-7-5 swimmer with metachronal waves oriented along the main axis ($k_\varphi=0$) 
		for various phase lags $\chi$ between successive rings and varying power stroke orientation $\theta_r$. 
		\textbf{a)} Average swimming velocity $<v_n>$. 
		\textbf{b)} Velocity fluctuations$\sqrt{<(v_n - <v_n>)^2>}$ around the average. 
		\textbf{c)} Rotational velocity $\Omega_n$ around the main body axis.
		\textbf{d)} Auto correlation function $<{\bf n}(t) \cdot {\bf n}(t+\tau)>_t$ for swimmers with 
		phase lag $\chi=-50^\circ$ (indicated by red stripe in (a-c)) and varying $k_\varphi$. 
		The dashed lines are fits to the auto-correlation function Eq.~(\ref{eq:correlations}) (see text).
	}
	\label{fig:vel_n474_oblique} 
\end{figure*}

Results for the swimming properties with oblique power-stroke direction are displayed in Fig.~\ref{fig:vel_n474_oblique}.
The results for the swimming velocity, velocity fluctuations, and rotational motion are qualitatively similar as
for the oblique metachronal wave. There is again a ``sinusoidal" dependence of the velocity $\langle v_n \rangle$ on 
the phase lag $\chi$, with a maximum for at $\chi=-50^\circ$, i.e. for an antiplectic metachronal coordination, see
Fig.~\ref{fig:vel_n474_oblique}a. However, there are also several pronounced qualitative and quantitative differences.
The velocity decreases with increasing $\theta_r$, because an increasing fraction of the beat is employed for
body rotation rather than forward propulsion. 
Velocity fluctuations $\sqrt{\langle(v_n - \langle v_n \rangle)^2 \rangle}$ peak for synchronously beating cilia for
all $\theta_r$, as all cilia beat in synchrony in this case (with $k_\varphi=0$), see Fig.~\ref{fig:vel_n474_oblique}b. 
The body rotation is much more pronounced for all
$\theta_r>0$, see Fig.~\ref{fig:vel_n474_oblique}c, compared to the case of oblique metachronal wave with 
$\theta_r=0$. Interestingly, the body rotation becomes very slow for $\chi\simeq 50^\circ$ for all $\theta_r>0$.
This minimum of spinning frequency $\Omega_n$ correlates with the minimum of propulsion velocity $v_n$, i.e. weak
propulsion is accompanied by slow spinning.
Correlation functions for fixed $\chi=-50^\circ$ and various $\theta_r$ are shown in Fig.~\ref{fig:vel_n474_oblique}d.

\begin{figure*}
	\centering
	\includegraphics[width=0.48\textwidth]{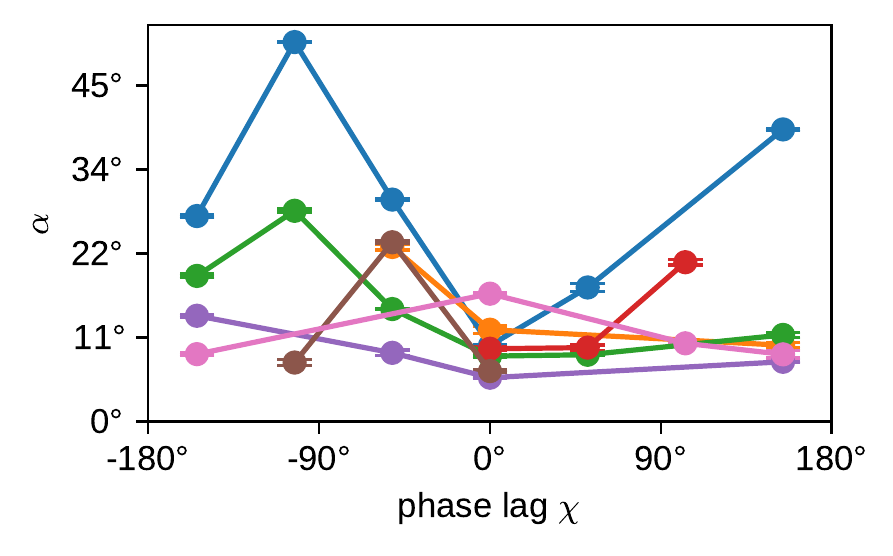} 
	\includegraphics[width=0.48\textwidth]{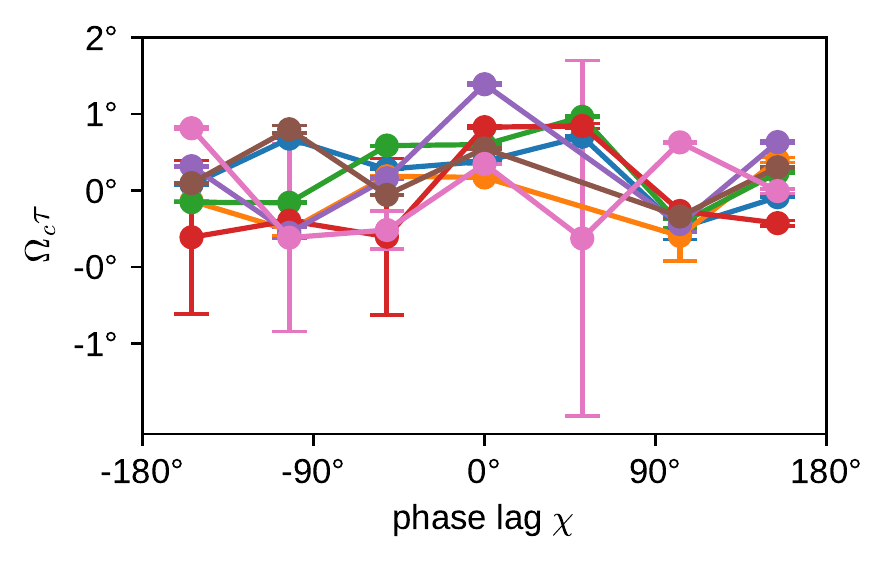} 
	\includegraphics[width=0.48\textwidth]{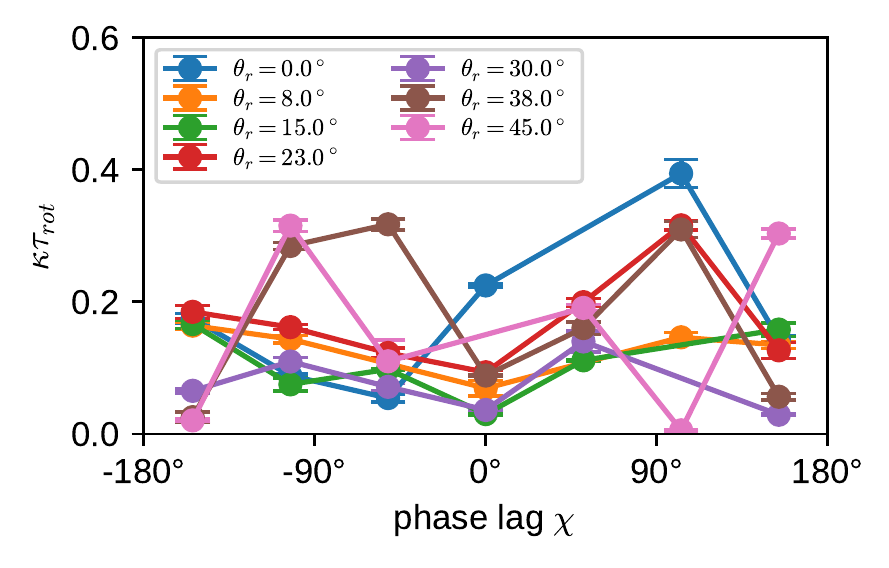} 
	\includegraphics[width=0.47\textwidth]{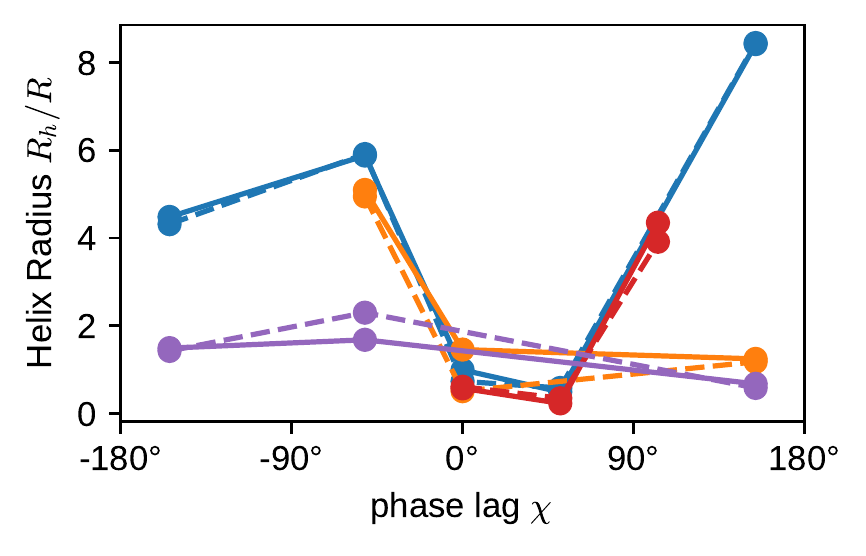}
	\caption{Swimming properties of a 5-7-5 swimmer with cilia beat in the longitudinal direction ($\theta_r=0$), 
		for various phase lags $\chi$ and latitudinal modes $k_\varphi$. 
		\textbf{a)} Alignment angle $\alpha$;
		\textbf{b)} Rotation frequency $\Omega_c$, normalized by the beat period $\tau$;
		\textbf{c)} Inverse correlation time $\kappa$, normalized by the rotational diffusion time $\tau_{rot} = 1/(2D_{rot})$,
		with $D_r=k_BT/(8\pi \eta R^3)$.
        \textbf{d)} Helix radius $R_h$ in units of the body radius $R$, as obtained from Eq.~(\ref{eq:helix_params}).
	}
	\label{fig:correlation_parameters_thetar}
\end{figure*}

Results for the fitted parameters of the helical motion and the decay time $1/\kappa$ in Eq.~(\ref{eq:correlations})
are displayed in Fig.~\ref{fig:correlation_parameters_thetar}. 
For most cases, the helix angles $\alpha$ are small, around $10^\circ$ to $25^\circ$, which implies that the constant 
term in Eq.~(\ref{eq:helix}) dominates over the oscillatory term, and the helices are very thin and elongated. 
The circling frequencies $\Omega_c$, displayed in Fig.~\ref{fig:correlation_parameters_thetar}(b), are generally
very small, which implies that cilia beat orientation with $\theta_r>0$ results more in spinning than in circling.
Scaled decorrelation times $1/\kappa \tau_{rot}$ are typically larger than $4$, which indicates a large hydrodynamic
radius, in agreement with the conclusions of Sec.~\ref{sec:longitudinal}. 

We can use again the expression (\ref{eq:helix_params}) to extract the helix radius $R_h$ 
from the data presented in Figs.~\ref{fig:vel_n474_oblique} and \ref{fig:correlation_parameters_thetar}. 
The results are shown in Fig.~\ref{fig:correlation_parameters_thetar}(d). They indicate that similarly as for the case 
$\theta_r=0$ displayed in Fig.~\ref{fig:correlation_parameters_theta_r0}(d), the helix radius $R_h$ is typically 
rather small, from nearly zero to just a few times the body radius $R$. The helix pitch is typically about a factor $10$
larger than the radius.

\subsection{Variation of Cilia Number}
\label{sec:number}

\begin{figure*}
	\centering
	\includegraphics[width=0.49\textwidth]{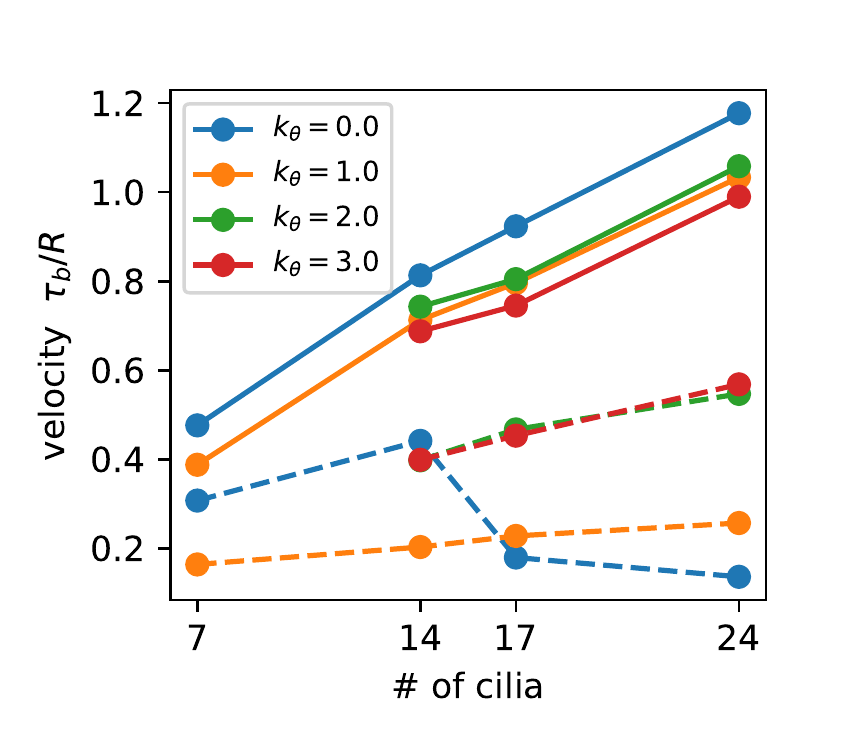} 
	\includegraphics[width=0.49\textwidth]{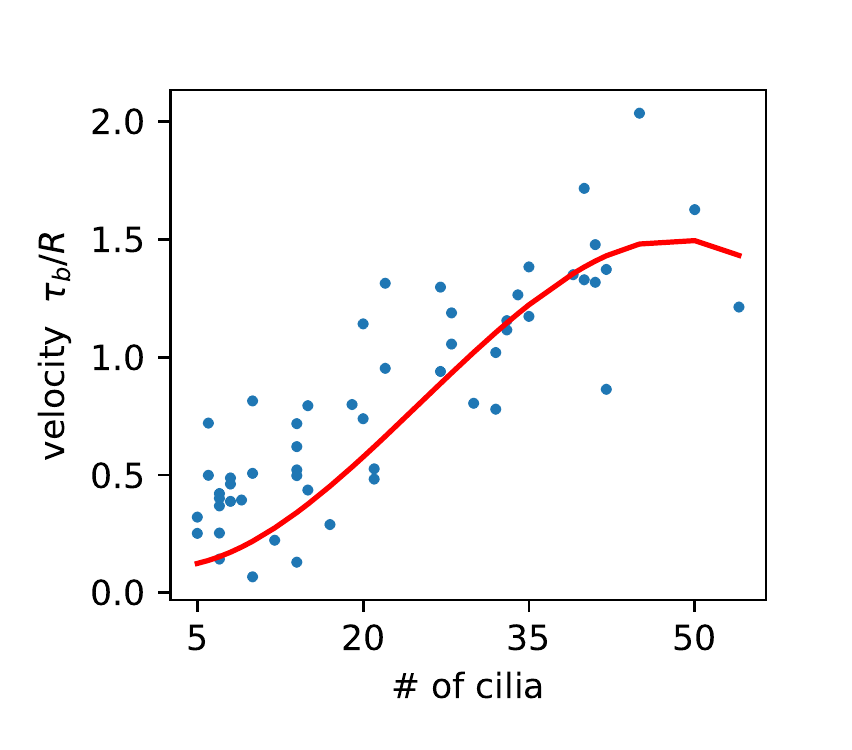} 
	\caption{Swimming velocity for microswimmers with various numbers of cilia.
		(a) With a predefined metachronal wave with various latitudinal wave 
		numbers $k_\varphi$, as indicated. Full (dashed) lines show that maximum (minimum) velocity obtained by 
		variation of the phase lag $\chi$.
        (b) With self-organized metachronal waves and random spatial arrangement. The red line is a spline fit to 
        different cilia arrangements. See also movie SM4 for an illustration of the beat and swimming behavior.}
	\label{fig:vel_number}
\end{figure*}

We have considered so far a spherical microswimmer with a fixed number $N=5+7+5=17$ cilia. It is now of 
course interesting to see how the swimming behavior, in particular the swimming velocity, depends on the 
number of cilia. For the swimmer with imposed metachronal wave and $\theta_r=0$, as described in Sec.~\ref{sec:longitudinal},
the results are shown in Fig.~\ref{fig:vel_number}(a), both for the maximum and minimum velocity obtained for various
phase lags $\chi$. In general, the maximum velocity is found to increase with
cilia number, consistent with the results of Ref.~\cite{ito19} for high cilia numbers. For all cilia numbers considered, 
the highest velocity is obtained for $k_\varphi=0$, i.e. for a wave direction along the main body axis, in agreement
with our arguments in Sec.~\ref{sec:longitudinal} above. In contrast, the variation of the minimum velocity on cilia
number is much less pronounced.

We want to compare this result with the case of a self-organized metachronal wave -- similar to what has been studied
in Ref.~\cite{Elgeti2013} for planar cilia arrays. The cilia are distributed randomly, but as homogeneously as possible
on the body surface. We consider several different distributions, so that the variance of the results for the same
cilia number can be taken as a measure for the sensitivity of the swimming velocity on the spatial distribution.
Results are shown in Fig.~\ref{fig:vel_number}(b). Again, the velocity increases with cilia number, but seems to level
off at about 40 cilia. This saturation is not too surprising, as with increasing cilia density, the effect of each
individual cilium on the propulsion diminishes due to the interaction with the neighbors. For $N$ cilia on the 
surface, the average distance $d$ between them is approximately determined by $d \simeq (4\pi R^2/N)^{1/2}$, 
which yields $d/L \simeq 1/6$ for $N=40$ and cilia length $L=3R$. 

The results for the self-organized metachronal wave on a spherical body can now be compared with those on a planar 
substrate as a function of $d/L$. For the planar case, the fluid transport velocity was found to scale as 
$v_{fluid} \sim (d/L)^{-\gamma}$ with $\gamma\simeq 1.4$ \cite{Elgeti2013}. This implies a dependence of the 
swimming velocity $v_{swim}$ on cilia number as $v_{swim} \sim N^{\gamma/2}$, i.e. a behavior somewhere between
linear and square root, which seems not inconsistent with the numerical results of Fig.~\ref{fig:vel_number}(b).
The simulation results of Ref.~\cite{ito19} for microswimmers with large cilia numbers
(in the range $N=20$ to $N=320$) also show a sublinear dependence of $v_{swim}$ on $N$, which a nearly
linear dependence for intermediate values of the phase lag $\chi$, and a strongly sublinear dependence
for $\chi=0$. 

It is also interesting to note that the results for cilia arrays on a planar substrate indicate that $d/L=1/6$
is in the regime where transport with metachronal coordination is much more efficient than synchronous 
beating \cite{Elgeti2013} -- in qualitative agreement with the results displayed in Fig.~\ref{fig:vel_number}(a).

\section{Summary and Conclusion}

The dynamics and motion of multi-ciliated microswimmers with a spherical body and a small number $N$, in the 
range $5 < N <50$, of long cilia, with length $L$ three times the body radius, has been investigated by
mesoscale hydrodynamics simulations. A metachronal wave is imposed for the cilia beat, for which the 
wave vector has both a longitudinal, $k_\theta$, and a latitudinal, $k_\varphi$ component. The dynamics
and motion is characterized by the swimming velocity $v_n$ along the main body axis, the variance of the 
velocity averaged over a full beat cycle, the spinning velocity $\Omega_n$ around the main body axis, as well
as the parameters of the helical trajectory, which are the circling velocity $\Omega_c$, the helix angle
$\alpha$, the helix radius $R_h$ and pitch $P_h$.

Our simulation results show that the microswimmer motion strongly depends on the latitudinal wave
number $k_\varphi$ and the longitudinal phase lag $\chi=k_\theta (\pi/4)$. We find, not unexpectedly,
that the microswimmers usually spin around their own axis, and swim on helical trajectories. 
It is important to notice that spinning and helical motion are not necessarily correlated, as a
spinning particle can move on a perfectly straight trajectory. However, chirality in the metachronal
beat pattern generically generates helical trajectories. In most cases, the helices are found to
be thin and stretched, i.e. the helix radius $R_h$ is about an order of magnitude smaller than the pitch.
An interesting result is also that the rotational diffusion of the microswimmer is significantly
smaller than the passive rotational diffusion of the body alone. This indicates that active 
contributions to rotational diffusion are small, and that the extended cilia make a pronounced
contribution to the hydrodynamic radius.

The swimming velocity $v_{swim}$ increases with the number $N$  of cilia on the body. Our simulation results 
indicate a slightly sublinear dependence on $N$. The comparison with the transport velocity of planar cilia 
arrays on the cilia separation predicts a dependence 
$v_{swim} \sim N^{0.7}$. Our simulation results for self-organized metachronal waves are not inconsistent 
with such a relation. 

Finally, it is interesting to note that already a relatively small number of about ten cilia,
beating with a phase lag in the form of a metachronal wave, are sufficient to generate 
a steady propulsion and smooth swimming motion -- in contrast to the strongly oscillatory motion of
Chlamydomonas \cite{guas10, geye13} with its two cilia beating in synchrony. 
Such a smooth swimming motion can be of significant benefit for marine microorganisms, because large 
disturbances can be exploited by predators to locate their prey \cite{kioe99}.

\ack
Support by the Deutsche Forschungsgemeinschaft (DFG) through the priority program on ``Microswimmers -- from 
single particle motion to collective behavior'' (SPP 1726) is gratefully acknowledged.
Computing time has been granted through JARA-HPC on the supercomputer JURECA \cite{jureca} 
at Forschungszentrum Jülich.

\appendix 
\section{Ciliary Beat}
\label{app:cil_beat}

\begin{figure}[ht]
  	\centering
  	\includegraphics{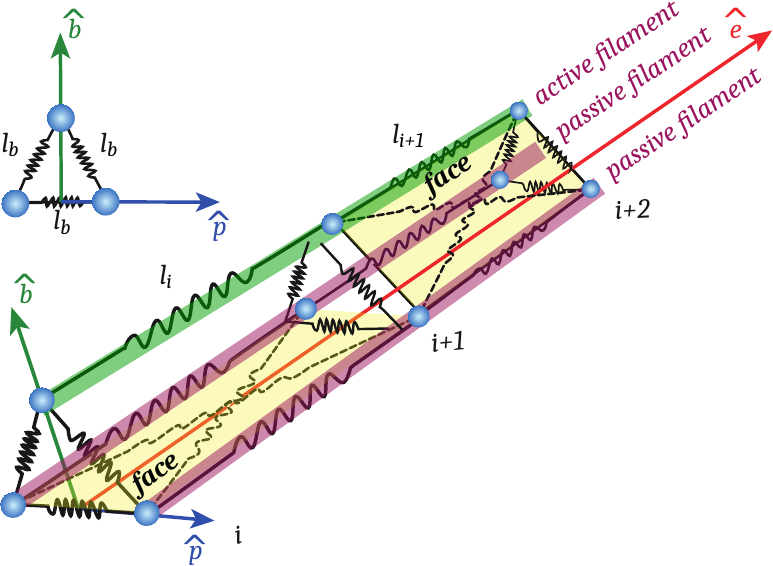} 
  	\caption{Sketch of the cilium model in the local reference frame $\hat{e},\hat{b},\hat{p}$- The cilium consists of one active rod (green) 
  		and two passive ones (red), forming an equilateral triangle. It's main axis is oriented along $\hat{e}$ and it beats in the $\hat{b}$ direction.  Each bead (blue) is connected by several springs (black) to its neighboring beads. Two diagonal springs connect the 
  		edges of each face. Not that they are only shown for the two yellow-marked faces to improve readability. Segments along $\hat{e}$ 
  		are labeled in ascending order, starting at $i=0$ and have a distance of $l_b=0.5~a$.}
  	\label{fig:cilium_constr}
\end{figure}
  
The cilium is considered as a semi-flexible filament, which is modeled by point particles (beads) that are interconnect by springs. 
Beads are arranged in three linear chains that form a bundle with cross-section of an equilateral triangle 
(see Fig.~\ref{fig:cilium_constr}). Each chain consist of $N_c=26$ individual beads equally spaced at a distance of $0.5~a$, 
where $a$ is the linear MPC collision-cell size. Springs with equilibrium length $l_b$ connect each bead to its neighboring 
beads along the line. Two diagonal connection on each face provide lateral stability (see Fig.~\ref{fig:cilium_constr}).   
The forces along the active chain are generated by changing the equilibrium length of the springs along the chain \cite{Elgeti2013}.  

The beat pattern is generated by a heuristic model for the time evolution of bond forces along the chain. In this description,
the beat can be controlled by a few key parameters, and allows adaptation to external flows. The beat is regulated by 
making the equilibrium curvature along the cilium depend on a pivot point $i_0$. 
Thus, the bond length along the active filament varies as
\begin{eqnarray}
  	l_i(t) &= l_b + \delta l_i(t) \nonumber \\
  	\delta l_i(t) &=  0.5 A \left(1 - \frac{i - n_0 - 1}{N_F -n_0 - 1} \right)^{2.5}  \nonumber \\
    	& \ \ \ \ \ \ \ \ \  \times \left( 1 - \frac{1}{i_0-i-1} \right) \ \ \ \ \textrm{for } i < i_0 -2 \nonumber \\
  	\delta l_i(t) &= - \frac{2A}{(i - i_0)^2+1}\  \ \ \ \   \textrm{for } i \geq i_0 - 2,
\end{eqnarray}
where $i$ is the segment number which varies from $n_0$ to $L/l_b$. The first three beads of each linear chain are passive and 
stay at the bond-length value $l_{i<n_0} = l_b$. Therefore, we set $n_0=3$.
  
The difference between power and recovery stroke depends on the pivot-point position $i_0$. During the power stroke, 
the pivot point is set to the first point of active beating $n_0$. It stays at this position, until the power stroke 
time $\tau_p$ has passed. Then, during the recovery stroke, the pivot-point $i_0$ moves along the cilium with a constant 
velocity $v_{rec}$ until the recovery-stroke time $\tau_r$ has passed. Together, this results in a beat as displayed 
in Fig.~\ref{fig:beat_pattern}, with a constant beat period $\tau_b = \tau_p + \tau_r$.

In addition to the global beat dynamic, the dynamics of the beat pattern is limited by the maximum energy $E_{max}$ 
each motor can inject into the system. 
If the bond-length change $\delta l_i$ compared to the current distance between the beads exceeds the maximal energy 
$E_{max}$ a motor can exert, the motor stall and $\delta l_i$ is reduced such that a maximum energy of $E_{max}$ is not 
exceeded. Due to the thermal noise in the system, the actual distance between two beads fluctuates as well. 
Although we do allow for ``helpful" fluctuations, we prohibit the ``obstructive" fluctuations by clipping the bond 
length at the previous value $\delta l_i(t+1) = \delta l(t)$.  
This update scheme ensures step-wise directional movement similar to an active Brownian ratchet.

The parameters of the cilia beat pattern are specified in Tab.~\ref{tab:parameters}. 

\renewcommand{\arraystretch}{1.5}
\begin{table}[htbp]
  	\centering
  	\begin{tabular}[5pt]{|c|c|c|c|}
  		\hline 
  		$l_b$ & $A$ & $\tau_p$ & $\tau_r$  \\ 
  		\hline 
  		$0.5~a$ & $0.17~a$ & $75~h$ & $150~h$ \\ 
  		\hline 
  	\end{tabular} \\[1ex]
  
  \centering
  	\begin{tabular}[5pt]{|c|c|c|c|}
	\hline 
	 $E_{max}$ & $L$ & $\gamma$ & $v_{rec}$ \\ 
	\hline 
	 $1~k_bT$ & $13~a$ & $20`000~k_bT/a^2$ & $0.19~l_b/h$ \\ 
	\hline 
    \end{tabular} 
  	\caption{Model parameters (given in MPC units of thermal energy $k_BT$, collision cell size $a$, and collision time $h$) 
  		used to generate the beat pattern of the cilium in the simulations.}
  	\label{tab:parameters}
\end{table}

\section{Parameter Extraction from Correlation Functions}

Due to the long time scales and the inherent noise of the auto-correlation function, it is difficult to extract parameters 
reliably. The signal-to-noise ratio is particularly high in the limit where the time scales $1/\Omega_c$ and $1/\kappa$ 
approach the total simulation time $T_s$. 

\begin{figure}[ht]
	\centering
	\includegraphics[width=\columnwidth]{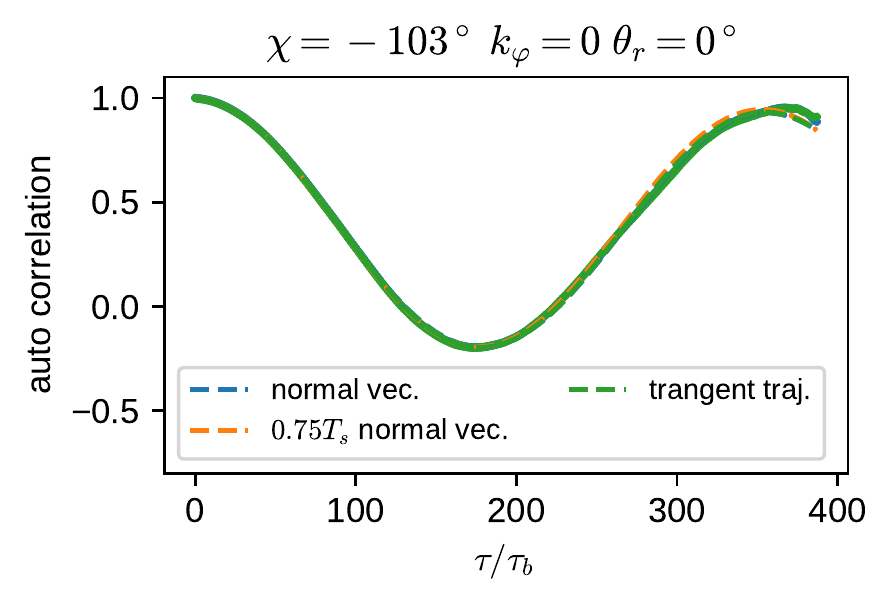} 
	\includegraphics[width=\columnwidth]{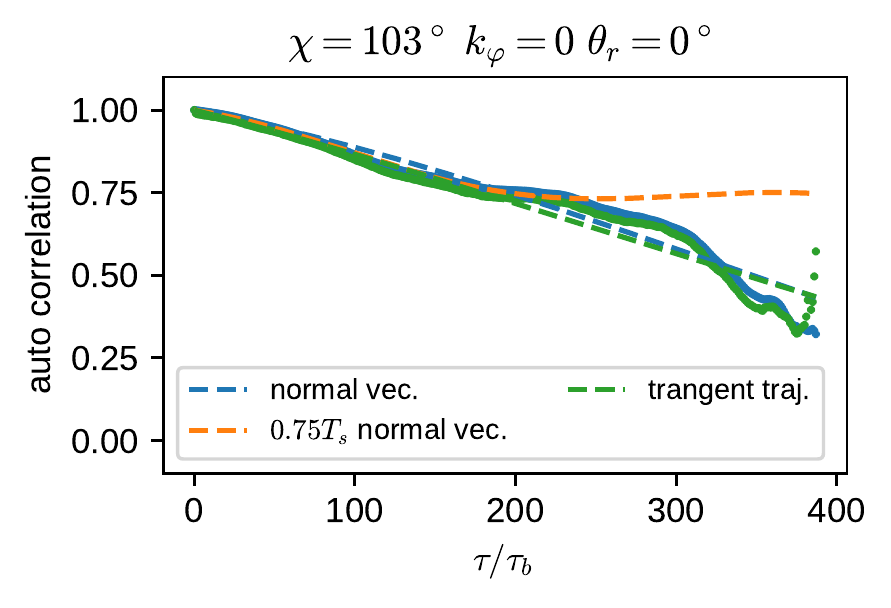} 
	\caption{Fit results of the temporal auto-correlation function which characterizes the helical trajectory. 
		Either the auto-correlation function of the unit normal vector ${\bf n}$ of the main body axis is employed,  
		or of the tangent vector along the smoothed center-of-mass trajectory, both with a temporal delay $\tau$.
	    Simulation data are shown as solid lines, fits to Eq.~(\ref{eq:correlations}) as dashed lines.}
	\label{fig:fit_quality}
\end{figure}

Due to the approximate rotational symmetry of the microswimmer, we assume that the tangent vector to its center-of-mass 
trajectory is along the main body axis $\vec{n}$.  This allows us to access the correlation functions, 
which determine the trajectory in two  different ways. Two examples for the quality of the fit results are shown 
in Fig.~\ref{fig:fit_quality}. In order to estimate the quality of the fits, we employ the difference between two 
different -- although related -- auto-correlation functions which characterize the helical swimming 
motion. On the one hand, we use the correlation function $\langle \vec{n}(t) \vec{n}(t+\tau) \rangle$ of the main body 
axis, as discussed in the main text; on the other hand, the correlate the tangent vectors of the smoothed center-of-mass 
trajectory with a window of $\tau$.

For a fixed total simulation time $T_s$, the error of the auto-correlation function increases proportional to 
$\sqrt{\tau/T_s}$, which is used as a weight factor for the fit. As a second error estimate, we employ the difference 
between a fit of the full trajectory with $\tau\le T_s$, and a fit for $\tau < 0.75~T_s$.
Parameters, where the fit does not converge, or where the difference between the parameter obtained from the full fit 
and a partial fit exceeds the error threshold displayed in Tab.~\ref{tab:thresholds}, are omitted and are not shown 
in the plots.

\renewcommand{\arraystretch}{1.5}
\begin{table}[htbp]
	\centering
	\begin{tabular}[5pt]{|c|c|}
		\hline 
		parameter & threshold  \\ 
		\hline 
		$\kappa$ & $0.2/\tau_{rot}$ \\ 
		$\Omega_c$ & $\pm 5^\circ/\tau_b$ \\ 
		$\alpha$ & $\pm 25^\circ$ \\
		\hline 
	\end{tabular} 
	\caption{Error thresholds for the fits of the auto-correlation functions.}
	\label{tab:thresholds}
\end{table}

Furthermore, the microswimmer can perform motions which more complex than a simple helix, e.g. a nutation-like motion. 
This is not captured by our analytic formula, Eq.~(\ref{eq:correlations}). We thus omit all results from fits with a too large
mean squared deviation.

\section*{References}
\bibliography{microswimmer,wan} 
\bibliographystyle{apsrev}

\end{document}